\documentclass{elsart}
\usepackage{epsfig,times,graphics}
\usepackage{amsmath,amsfonts,amssymb}
\usepackage{cite,graphicx,subfigure,url}
\usepackage{setspace}

\newcommand{\SF}{ {\mathcal F} }

\newcommand {\tE}{\textrm{E}}

\newtheorem{definition}{{\bf Definition}}
\newtheorem{theorem}{{\bf Theorem}}
\newtheorem{lemma}{\hspace{-0.15in}{\bf Lemma}}
\newtheorem{proposition}{\hspace{-0.15in}{\bf Proposition}}

\def\done{\hspace*{\fill} \rule{1.8mm}{2.5mm} }

\begin{document}

\begin{frontmatter}

\title{{\bf On Effectiveness of Backlog Bounds Using Stochastic Network Calculus in 802.11}}

\author{Yue Wang (corresponding author)}
\address{
      School of Information\\
      Central University of Finance and Economics\\
      Address: 39 South College Road, Haidian District,Beijing,China 100081\\
}

\begin{abstract}
Network calculus is a powerful methodology of characterizing
queueing processes and has wide applications, but few works on
applying it to 802.11 by far. In this paper, we take one of the
first steps to analyze the backlog bounds of an 802.11 wireless LAN
using stochastic network calculus. In particular, we want to address
its effectiveness on bounding backlogs. We model a wireless node as
a single server with impairment service based on two best-known
models in stochastic network calculus: Jiang's and Ciucu's.
Interestingly, we find that the two models can derive equivalent
stochastic service curves and backlog bounds in our studied case. We
prove that the network-calculus bounds imply stable backlogs as long
as the average rate of traffic arrival is less than that of service,
indicating the theoretical effectiveness of stochastic network
calculus in bounding backlogs. From A. Kumar's 802.11 model, we
derive the concrete stochastic service curve of an 802.11 node and
its backlog bounds. We compare the derived bounds with ns-2
simulations and find that the former are very loose and we discuss
the reasons. And we show that the martingale and independent case
analysis techniques can improve the bounds significantly. Our work
offers a good reference to applying stochastic network calculus to
practical scenarios.
\end{abstract}


\end{frontmatter}

\section{{\bf Introduction}}\label{sec:introduction}

Network calculus provides an elegant way to characterize traffic and
service processes of network and communication systems. Unlike
traditional queueing theory in which one has to make strong
assumptions on arrival or service processes (e.g., Poission arrival
process, exponential service distribution, etc) so as to derive
closed-form solutions in queueing networks\cite{queueing_book},
network calculus allows general arrival and service processes.
Instead of getting exact solutions, one derives network backlog and
delay bounds by network calculus. Deterministic network calculus is
mature in theory
\cite{netcal_cruz_1}\cite{netcal_cruz_2}\cite{boudec_book}\cite{chang_book}.
However, most traffic and service processes are stochastic and
deterministic network calculus is often not applicable to them.
Therefore, stochastic network calculus was proposed to deal with
stochastic arrival and service processes
\cite{chang_book}-\cite{ciucu_martingale}\cite{app_80211_1}. 

Numerous applications of it have been found in communication
networks and even in management science, and we cite some of them
\cite{app_mbac}-\cite{app_mbs}. However, few works have been made on
applying it to multiple access communication networks such as 802.11
Wireless LANs\cite{app_nonasymp}\cite{app_80211_1}. In the paper, we
take one of the first steps to apply stochastic network calculus to
an 802.11 wireless LAN (WLAN). In particular, we want to address the
effectiveness of stochastic network calculus on bounding backlogs in
802.11, with the following sub-problems:
\begin{itemize}
\item Under what condition can we derive stable backlogs using network calculus?
\item How to derive the \emph{concrete} stochastic service curve an 802.11 node?
\item Are the derived backlog bounds tight compared with ns-2 simulations? And how to improve them?
\end{itemize}

We model a wireless node as a single server with impairment service
based on two best-known models in stochastic network calculus:
Jiang's\cite{jiang_snc3} and Ciucu's\cite{ciucu_snc}. And we make
the following \emph{new} contributions on this topic:

\begin{itemize}
\item
We compare Jiang's and Ciucu's model and find that they can derive
equivalent stochastic service curves and backlog bounds in our
studied case.
\item We prove that the network-calculus backlog bounds imply stable backlog as long as the average rate of
traffic arrival is less than that of service, indicating that
stochastic network calculus is effective in bounding backlogs
\emph{theoretically}.
\item From A. Kumar's 802.11 model, we derive the \emph{concrete} stochastic service curve of an 802.11 node
\cite{kumar_802_11} and give the numerical computation methods. From
the service curve we then derive backlog bounds.
\item We observe the derived bounds are loose when compared with ns-2 simulations.
However, the martingale and independent case analysis techniques can
improve the bounds significantly.
\end{itemize}

Note that when we prove a statement in this paper, we call it
\emph{Propositions} to differentiate the existing theorems in the
literature (see Proposition
\ref{prop:vb_improve}-\ref{prop:model_equiv}).

This paper is organized as follows. In Section~\ref{sec:snc}, we
give a brief overview of stochastic network calculus. In particular,
we present the classic models of Jiang's and Ciucu's and we also
discuss the martingale and independent case analysis techniques.
In Section~\ref{sec:model}, we present the network calculus model of
a wireless node based on Jiang's and Ciucu's model. We compare the
two models and find that they are equivalent in deriving  stochastic
service curves and backlog bounds in our studied case. We also prove
the stability condition by the theory of stochastic network calculus
in this section.
In Section~\ref{sec:802_11}, we derive the backlog bounds of an
802.11 node and the critical part is to derive its \emph{concrete}
stochastic service curve.
In Section~\ref{sec:simulation}, we compare the derived backlog
bounds with ns-2 simulation results under Poisson traffic arrivals.
In particular, we show that the martingale and independent case
analysis techniques can improve the bounds significantly.
In Section~\ref{sec:related}, we give related works and highlight
our contributions. Finally, Section~\ref{sec:conclusion} concludes
the paper and points out some future works.

\section{{\bf Stochastic Network Calculus}}\label{sec:snc}

In this section, we first review basic terms of network calculus and
then cite some results of the stochastic network calculus theory
used in our paper. Jiang classified stochastic arrival curves as the
types of \emph{ta (traffic amount centric)}, \emph{vb (virtual
backlog centric)} and \emph{mb (max virtual backlog centric)}, and
classified stochastic service curve as \emph{ws (weak stochastic)}
and \emph{sc (stochastic)}. In this paper, we adopt ta and mb
arrival curves and the ws service curve, as currently they provides
tightest backlog bounds\footnote{As recently known by the network
calculus community the bounding probability of a mb arrival curve is
1 for linear arrival curve functions, making its usage restrictive.
So does the sc service curve as it is often derived from an
impairment process with the mb arrival curve.}. Note that we just
say \emph{"stochastic service curve"} in our paper which means the
\emph{ws} one.

\ \
\subsection{{\bf Basic Terms of Network Calculus}} \label{subsec:terms}

\begin{figure}[htb]
  \centering
  \includegraphics[width=0.8\textwidth]{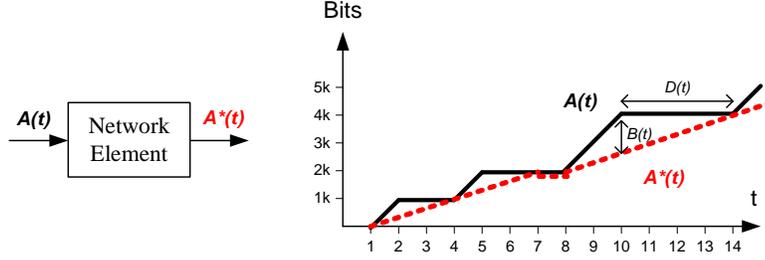}\\
  \caption{Illustration of $A(t)$, $A^*(t)$, $B(t)$ and $D(t)$}\label{fig:curves_eg}
\end{figure}

We consider a discrete time system where time is slotted ($t = 0, 1,
2, ...$). A process is a function of time $t$. By default, we use
$A(t)$ to denote the \emph{arrival process} to a network element
with $A(0)=0$. $A(t)$ is the total amount of traffic arrived to this
network element up to time $t$. We use $A^*(t)$ to denote the
\emph{departure process} of the network element with $A^*(0)=0$.
$A^*(t)$ is the total amount of traffic departed from the network
element up to time $t$. Let $\SF$ ($\bar{\SF}$) represents the set
of non-negative wide-sense increasing (decreasing) functions.
Clearly, $A(t)\in \SF$ and $A^*(t)\in \SF$. For any process, say
$A(t)$, we define $A(s,t) \equiv A(t)-A(s)$, for $s \leq t$. We
define the backlog of the network element at time $t$ by
\begin{eqnarray}\label{eq:backlog}
B(t)=A(t)-A^*(t),
\end{eqnarray}
and the delay of the network element at $t$ by
\begin{eqnarray}\label{eq:delay}
D(t)=\inf\{d: A(t)\leq A^*(t+d)\}.
\end{eqnarray}
Fig.~\ref{fig:curves_eg} illustrates an example of $A(t)$ and
$A^*(t)$ with $B(t)$ and $D(t)$ at $t=10$.

In deterministic network calculus, $A(t)$ can be upper-bounded by an
arrival curve. That is, for all $0 \leq s \leq t$, we have
\[A(s, t) \leq \alpha(t-s),\]
where $\alpha(t)$ is called the \emph{arrival curve} of $A(t)$.

We say, \emph{busy period} is a time period during which the backlog
in the network element is always nonzero. For any busy period $(t_0,
t]$, suppose we have
\[A^*(t) - A^*(t_0) \geq \beta(t-t_0),\]
if the network element provides a guaranteed service lower-bounded
by $\beta(t-t_0)$ during the busy period. We can let $t_0$ be the
beginning of the busy period, that is, the backlog at $t_0$ is zero
or $A^*(t_0)=A(t_0)$. Therefore,
\[A^*(t) - A(t_0) \geq \beta(t-t_0).\]

The above equation infers $A^*(t) \geq \inf_{0\leq s \leq
t}{[A(s)+\beta(t-s)]}$, which can be written as
\begin{eqnarray}
A^*(t) \geq A\otimes\beta(t),
\end{eqnarray}
where $\otimes$ is called the operator of \emph{min-plus
convolution} and $\beta(t)$ is called the \emph{service curve} of
the network element.

\ \
\subsection{{\bf Stochastic Network Calculus Theory}}\label{subsec:snc_theory}

We consider a server $S$ (i.e. the network element) fed with a flow
$A$. In practice, $A$'s traffic and $S$'s service are often
stochastic, which can not be hard bounded by some curves. That is,
they can violate the curves but with certain probabilities (we call
it \emph{bounding function} here). The theory of stochastic network
calculus can get probabilistic bounds for backlogs and delays of the
server, suppose we can characterize $A$ by a stochastic arrival
curve and $S$ by a stochastic service curve.

In this section, we just consider the derivation of backlog bounds
as delay bounds are quite similar to the former. We first give some
definitions. Then we cite some results in Jiang's and Ciucu's
models\cite{jiang_snc3}\cite{ciucu_snc} and the construction .
Lastly, we make a brief discussion on them.

\ \
\subsubsection{{\bf Definitions}}\label{subsec:defs}

\begin{definition}[ta stochastic arrival curve]\label{def:ta_ac}

 A flow is said to have a
\emph{ta (traffic-amount-centric) stochastic arrival curve} $\alpha
\in \SF$ with bounding function $f \in \bar{\SF}$, denoted by $A
\sim_{ta} <f, \alpha>$, if for all $s, t \geq 0$($s\leq t$) and all
$x \geq 0$, there holds
\begin{eqnarray}
P\{A(s, t) - \alpha(t-s)>x\} \leq f(x).
\end{eqnarray}
\end{definition}

\begin{definition}[vb stochastic arrival curve]\label{def:ac}
 A flow is said to have a
\emph{vb (virtual-backlog-centric) stochastic arrival curve} $\alpha
\in \SF$ with bounding function $f \in \bar{\SF}$, denoted by $A
\sim_{vb} <f, \alpha>$, if for all $t \geq 0$ and all $x \geq 0$,
there holds
\begin{eqnarray}
P\{\sup_{0 \leq s \leq t}[A(s, t) - \alpha(t-s)]>x\} \leq f(x).
\end{eqnarray}
\end{definition}
We can see that $A \sim_{vb} <f, \alpha>$ implies $A \sim_{ta} <f,
\alpha>$, since $P\{A(s, t) - \alpha(t-s)>x\} \leq P\{\sup_{0 \leq s
\leq t}[A(s, t) - \alpha(t-s)]>x\}$.

\begin{definition}[Stochastic Service Curve]\label{def:sc}
  A server $S$ is said to provide a
\emph{(weak) stochastic service curve} $\beta \in \SF$ with bounding
function $g \in \bar{\SF}$, denoted by $S \sim_{ws} <g, \beta>$ (or
just $S \sim <g, \beta>$), if for all $t \geq 0$ and all $x \geq 0$,
there holds
\begin{eqnarray}
P\{A \otimes \beta(t) - A^*(t) > x\} \leq g(x).
\end{eqnarray}
\end{definition}

\begin{definition}[Leftover Service]\label{def:leftover}
Consider a server $S$ provides the ideal service curve
$\hat{\beta}(t)$ with the impairment process $I$ to a flow. Then,
during any backlogged period $(s, t]$, the output flow $A^*(s, t)$
from the server satisfies
\begin{eqnarray}
A^*(s,t) \geq \hat{\beta}(t-s)-I(s,t).
\end{eqnarray}
$\hat{\beta}(t)-I(t)$ is the \emph{leftover service} received by the
given flow.
\end{definition}

The definition of leftover service (also called \emph{stochastic
strict server} in \cite{jiang_snc3}) can be applied to many
scenarios such as cross traffic and wireless channels.

\begin{definition}[$\theta$-MER] \label{def:theta-MER} A process $A$'s \emph{minimum envelope rate
 with respect to $\theta$ ($\theta$-MER)}, denoted by
$\rho^*(\theta)$, is defined as follows:
\begin{eqnarray}
\rho^*(\theta) = \lim\sup_{t\rightarrow \infty}\frac{1}{\theta
t}\sup_{s\geq 0}\log{\tE e^{\theta A(s,s+t)}}.
\end{eqnarray}
We say that $A$ has an \emph{envelope rate with respect to $\theta$}
($\theta$-ER), denoted by $\rho(\theta)$,  if $\rho(\theta) \geq
\rho^*(\theta)$.
\end{definition}

\begin{definition}[$(\sigma(\theta), \rho(\theta))$-upper constrained]
\label{def:sigma_rho} A process $A$ is said to be $(\sigma(\theta),
\rho(\theta))$\emph{-upper constrained} for some $\theta > 0$, if
for all $0 \leq s \leq t$, we have
\begin{eqnarray}
\frac{1}{\theta} \log{\tE e^{\theta A(s,t)}} \leq \rho(\theta)(t-s)
+ \sigma(\theta).
\end{eqnarray}
\end{definition}

We can derive stochastic arrival and service curves from the
$(\sigma(\theta), \rho(\theta))$-upper constrained characterization
(Section~\ref{subsec:computation}.

\begin{definition}[Average Rate]\label{def:avg_rate} The average rate of a process
$A$, denoted by $a_A$, is defined as
\begin{eqnarray}
a_A = \lim_{t\rightarrow \infty}\sup_{s\geq 0}\frac{\tE
A(s,s+t)}{t}.
\end{eqnarray}
\end{definition}

\begin{definition}[Stable Backlog Bound]\label{def:stable}
The backlog $B(t)$ is \emph{stable}, if for all $t$,
\begin{eqnarray}
\tE B(t) < C < \infty,
\end{eqnarray}
where $C$ is a finite constant value. We say that \emph{the backlog
bounds are stable} if they can derive stable backlogs.
\end{definition}

\ \
\subsubsection{{\bf Jiang's Model}}\label{subsec:jiang}

Jiang's model deals with \emph{vb arrival curves} and stochastic
service curves. We have the following theorems for leftover service
curves and backlog bounds.

\begin{theorem}[Jiang's Leftover Stochastic Service
Curve]\label{theo:jiang_leftover}
 Suppose a server $S$ providing the ideal
service curve $\hat{\beta}(t)$ with the impairment process $I$. If
$I$ has a \emph{vb stochastic arrival curve}, i.e., $I \sim_{vb} <g,
\gamma>$, then the server provides the flow the leftover stochastic
service curve $S \sim <g, \beta>$ and
\begin{eqnarray}
\beta(t) = \hat{\beta}(t) - \gamma(t).
\end{eqnarray}
\end{theorem}

\begin{theorem} [Jiang's Backlog Bounds] \label{theo:jiang_backlog}
If the flow $A$ has a \emph{vb stochastic arrival curve} $A
\sim_{vb} <f, \alpha>$ and the server $S$ provides a stochastic
service curve $S \sim <g, \beta>$ to the flow, then the
\emph{backlog} $B(t)$ of the flow in the server at time $t$
satisfies:
\begin{eqnarray}
P\{B(t) > x+\sup_{s \geq 0}{[\alpha(s)-\beta(s)]}\} \leq f \otimes
g(x),
\end{eqnarray}
for all $t \geq 0$ and all $x \geq 0$.
\end{theorem}

\ \
\subsubsection{{\bf Ciucu's Model}}\label{subsec:ciucu}

Ciucu's model deals with \emph{ta arrival curves} and stochastic
service curves.

In fact, we can derive vb arrival curves from ta arrival curves by
introducing the function $\delta(t)=\delta \cdot t$ ($\delta$ is an
adjustable constant). The following lemma states this.

\begin{lemma}[ta to vb Arrival Curves]\label{lemma:ta_vb}
Suppose $A$ is a ta stochastic arrival curve, $A \sim_{ta}
<f,\alpha>$, then $A \sim_{vb} <\tilde{f},\alpha_{\delta>}$ with
$\alpha_{\delta}(t) \equiv \alpha(t)+\delta t$ and its bounding
function $\tilde{f}(x,\delta) = \sum_{k=0}^{\infty}{f(x+k \delta)}$
(suppose the sum is finite).
\end{lemma}

The derivations are as follows.
\begin{eqnarray}
&&P\{\sup_{0\leq s\leq t}{[A(s,t)-\alpha_{\delta}(t-s)]>x}\}  \nonumber \\
&&\leq \sum_{s=0}^t {P\{A(s,t)-\alpha_{\delta}(t-s)>x\}} \nonumber\\
&&= \sum_{s=0}^t {P\{A(s,t)-\alpha(t-s)>x+\delta(t-s)\}}  \nonumber \\
&&\leq \sum_{s=0}^t {f(x+\delta(t-s))} \leq
\sum_{k=0}^{\infty}{f(x+k \delta)}. \label{eq:ta_vb}
\end{eqnarray}

We have the following theorems for leftover service curves and
backlog bounds in Ciucu's model. Actually, we can derive these
results by first converting ta arrival curves to vb ones and then
applying Jiang's theorems.

\begin{theorem}[Ciucu's Leftover Stochastic Service Curve]\label{theo:ciucu_leftover}
Suppose a server $S$ providing the ideal service curve
$\hat{\beta}(t)$ with the impairment process $I$. If $I$ has a
\emph{ta stochastic arrival curve}, i.e., $I \sim_{ta} <g, \gamma>$,
then the server provides the flow the leftover stochastic service
curve $S \sim <\tilde{g}, \beta>$ and
\begin{eqnarray}
\beta(t) = \hat{\beta}(t) - \gamma_{\delta}(t),
\end{eqnarray}
 where $\gamma_{\delta}(t) \equiv \gamma(t)+\delta t$ and
$\tilde{g}(x,\delta) \equiv \sum_{k=0}^{\infty}{g(x+k\delta)}$  by
definition.
\end{theorem}

\begin{theorem} [Ciucu's Backlog Bounds] \label{theo:ciucu_backlog}
 If the flow $A$ has a \emph{ta stochastic
arrival curve} $A \sim_{ta} <f, \alpha>$ and the server $S$ provides
a stochastic service curve $S \sim <g, \beta>$ to the flow, then the
\emph{backlog} $B(t)$ of the flow in the server satisfies: for all
$t \geq 0$ and all $x \geq 0$,
\begin{eqnarray}
P\{B(t) > x+\sup_{s \geq 0}{[\alpha_{\delta}(s)-\beta(s)]}\} \leq
\tilde{f} \otimes g(x),
\end{eqnarray}
where $\alpha_{\delta}(t)\equiv \alpha(t)+\delta t$ and
$\tilde{f}(x,\delta) = \sum_{k=0}^{\infty}{f(x+k\delta})$ by
definition.
\end{theorem}

Note that Ciucu's can deal with ta arrival curves while Jiang's can
not, by introducing a $\delta > 0$ to trades smaller service for
finite bounding functions.

\ \ \subsection{{\bf Computation of Stochastic Arrival/Service
Curves}}\label{subsec:computation}

We will show in this subsection how to calculate stochastic arrival
and service curves from the $(\sigma(\theta),
\rho(\theta))$\emph{-upper constrained} characterization
\cite{chang_book}.

\begin{theorem}[Arrival Curves of $(\sigma(\theta),
\rho(\theta))$-\emph{upper constrained}]\label{theo:ac_comput}
Suppose $A(t)$ is $(\sigma(\theta),\rho(\theta))$-upper constrained,
then it has a ta stochastic arrival curve $A \sim_{ta} <f, \alpha>$,
where
\begin{eqnarray}\label{eq:ta_comput}
\alpha(t) &=& r \cdot t\nonumber\\
f(x) &=& e^{\theta\sigma(\theta)}\cdot e^{-\theta x},
\end{eqnarray}
for any $r \geq \rho(\theta)$ and $x \geq 0$. And $A$ has a vb
stochastic arrival curve $A \sim_{vb} <f, \alpha>$, where
\begin{eqnarray}\label{eq:vb_comput}
\alpha(t) &=& r \cdot t\nonumber\\
f(x) &=&
\frac{e^{\theta\sigma(\theta)}}{1-e^{\theta(\rho(\theta)-r)}}\cdot
e^{-\theta x},
\end{eqnarray}
for any $r > \rho(\theta)$ and $x \geq 0$.
\end{theorem}

Note that we have $r \geq \rho(\theta)$ in ta and $r > \rho(\theta)$
in vb. And Eq.(\ref{eq:vb_comput}) applies Boole's inequality to the
bound functions $f(x)$ which are loose in general.

How to derive stochastic service curves? If we can model the server
$S$ with the ideal service curve $\hat{\beta}$ with the impairment
process $I(t)$, we can first characterize $I(t)$ by vb (ta) arrival
curves, and then we use Theorem~\ref{theo:jiang_leftover}
(Theorem~\ref{theo:ciucu_leftover}) to get its stochastic service
curves.

The following theorem states the relation between $\theta$-ER and
$(\sigma(\theta), \rho(\theta))$-upper constrained. We will use it
in proving the stability condition of backlog bounds in
Section~\ref{subsec:stability}.

\begin{theorem}[$\theta$-ER vs $(\sigma(\theta),\rho(\theta))$-upper
constrained]\label{theo:theta-ER}
 If the process
$A(t)$ has a $\theta$-envelop rate ($\theta$-ER)
$\rho(\theta)<\infty$, then for every $\epsilon > 0$ there exists
$\sigma_\epsilon(\theta) < \infty$ so that $A$ is
$(\sigma_\epsilon(\theta), \rho(\theta)+\epsilon)$-upper
constrained.
\end{theorem}

\ \ \subsection{{\bf Improvement on Bounds }} \label{subsec:improve}
There are two ways of improving bounds in current literature. One
way is to apply independent case analysis. The other way is to
improve the bounding functions of stochastic service curves for
time-independent arrivals.

The first way says that suppose the impairment process of the server
$S$ is independent from the traffic arrival process, we can derive
tighter backlog bounds using independent probability analysis.

\begin{theorem}[Backlog Bounds under Independent Cases]\label{theo:independent}
Suppose the server $S$ provides the flow (satisfying $A \sim_{vb}
<f,\alpha>$) the ideal service curve $\hat{\beta}(t)$ with the
impairment process $I \sim_{vb} <g,\gamma>$ (thus $S \sim <g,\beta>$
where $\beta(t)=\hat{\beta(t)}-\gamma(t)$). Suppose \emph{$A$ and
$I$ are independent}, we have
\begin{eqnarray}
P\{B(t) > \sup_{s\geq 0}[\alpha(s)-\beta(s)]+x\} \nonumber\\
\leq \sum_{k=0}^x (\bar{g}(k)-\bar{g}(k-1))\bar{f}(x-k)
\end{eqnarray}
where $\bar{f}(x)=1-f(x)$, $\bar{g}(x)=1-g(x)$, and we set
$\bar{g}(-1)=0$.
\end{theorem}

This theorem of independent case analysis can be applied to Ciucu's
model. However, we first need to convert the ta arrival curves to
the vb ones with new bounding functions $\hat{f}(x,\delta)$ and
$\hat{g}(x,\delta)$ by Lemma~\ref{lemma:ta_vb}. Then we apply the
above theorem by plugging in $\hat{f}$ and $\hat{g}$.

Another way of tightening backlog bounds is to derive tighter
bounding functions of stochastic arrival and service curves. Ciucu
first proposed to use martingale to tighten the bounds for $M/M/1$
and $M/D/1$ queues\cite{ciucu_martingale}. In the following
proposition, we provide a more general result following his idea.
The proof is given in Appendix-A.

\begin{proposition}[vb Arrival Curves of Time-Independent
Process]\label{prop:vb_improve} Suppose $A(t)$ is
$(\sigma(\theta),\rho(\theta))$-upper constrained. On condition that
$a(t)\equiv A(t)-A(t-1)$ is independent of each $t$, it has a vb
stochastic arrival curve $A \sim_{vb} <f, \alpha>$, where

\begin{eqnarray}\label{eq:vb_improve}
\alpha(t) &=& r \cdot t\nonumber\\
f(x) &=& e^{-\theta x},
\end{eqnarray}

for any $r \geq \rho(\theta)+\sigma(\theta)$ and $x \geq 0$.
\end{proposition}

\ \ \subsection{{\bf Discussion on Jiang's and Ciucu's
Models}}\label{subsec:discussion}

The key difference between Jiang's and Ciucu's models is: Jiang use
vb traffic arrival curves while Ciucu uses ta ones variant
stochastic service curves. Which one can derive tighter backlog
bounds?

In general, ta arrival curves provide tighter bounding functions
than vb. Actually, $A \sim_{vb} <f, \alpha>$ implies $A \sim_{ta}
<f, \alpha>$ and the inverse is not true generally. In particular,
the bounding function of ta is tighter than that of vb (especially
when $r$ is close to $\rho(\theta)$) in
Theorem~\ref{theo:ac_comput}. But one can not conclude that Ciucu's
model is always better than Jiang's, as it has looser bounding
functions for leftover service curves and backlog bounds (see
Theorem~\ref{theo:ciucu_leftover} and
Theorem~\ref{theo:ciucu_backlog}). The situation becomes even more
uncertain when consider time-independent processes and independent
$A$ and $I$. Interestingly, we find that two models can derive
equivalent stochastic service curves and backlog bounds in our
studied case (Section~\ref{subsec:equiv}).

\section{{\bf A Wireless Node's Network Calculus Model}}\label{sec:model}
In this section, we model a general wireless node by stochastic
network calculus. In general, we can define one time slot ($t=1$) to
be any small duration and measure traffic amount in any unit (e.g.
bits, bytes or packets).

We consider a wireless node. Let $A(t)$ denote the traffic arrived
at the node from the application layer. We assume $A$ is
$(\sigma_{A}(\theta_1), \rho_{A}(\theta_1))$-upper constrained,
which is a right assumption for many cases.

We model the service of a wireless node as an ideal server curve
with an impairment process. In fact, the channel is shared by the
other node in a WLAN and transmission errors occur due to path loss
, fading and collisons, which contribute to the impairment process.
Let the channel capacity be $c$ traffic units per slot. The
departure process $A^*(s, t) = \hat{\beta}(s, t) - I(s, t)$ during
any backlogged period $[s,t]$, where $\hat{\beta}(t) = c\cdot t$ is
the ideal service curve and $I$ is the impairment process. Since
$I(s,t) \leq c\cdot(t-s)$, there exist $\sigma_{I}(\theta_2)$ and
$\rho_{I}(\theta_2)$ so that $I$ is $(\sigma_{I}(\theta_2),
\rho_{I}(\theta_2))$-upper constrained.

In here, $\theta_1$ and $\theta_2$ are adjustable parameters. We
will show Section~\ref{sec:802_11} how to calculate
$\rho_A(\theta_1)$, $\sigma_A(\theta_1)$, $\rho_I(\theta_2)$ and
$\sigma_I(\theta_2)$ for an 802.11 node.

\subsection{Jiang's Backlog Bounds}\label{subsec:jiang_bound}

Because $A$ is $(\sigma_{A}(\theta_1), \rho_{A}(\theta_1))$-upper
constrained, by Theorem~\ref{theo:ac_comput}, $A \sim_{vb}
<f,\alpha>$ where
\begin{eqnarray}\label{eq:jiang_A}
\alpha(t) &=& r_A \cdot t\nonumber\\
f(x) &=&
\frac{e^{\theta_1\sigma_{A}(\theta_1)}}{1-e^{\theta_1(\rho_{A}(\theta_1)-r_A)}}\cdot
e^{-\theta_1 x},
\end{eqnarray}
for any $r_A > \rho_{A}(\theta_1)$.

In the same way, $I \sim_{vb} <g,\gamma>$ where
\begin{eqnarray}\label{eq:jiang_I}
\gamma(t) &=& r_I \cdot t\nonumber\\
g(x) &=&
\frac{e^{\theta_2\sigma_{I}(\theta_2)}}{1-e^{\theta_2(\rho_{I}(\theta_2)-r_I)}}\cdot
e^{-\theta_2 x},
\end{eqnarray}
for any $r_I > \rho_{I}(\theta_2)$.

By Theorem~\ref{theo:jiang_leftover}, the node provides a stochastic
service curve $S \sim <g, \beta>$, where
\begin{eqnarray}\label{eq:jiang_S}
\beta(t) &=& (c-r_I) \cdot t,
\end{eqnarray}
for any $c > r_I$.

Finally, by Theorem~\ref{theo:jiang_backlog}, we must let $\alpha(t)
\leq \beta(t)$, i.e., $r_A \leq c - r_I$, in order to get meaningful
backlog bounds which are $P\{B(t)>x\} \leq f\otimes g(x)$.

We note that $f(x)$ ($g(x)$) is the decreasing function of $r_A$
($r_I$). Considering the above conditions, we get the following
optimal backlog bounds,
\begin{eqnarray}\label{eq:jiang_backlog}
&&P\{B(t) > x\} \leq \min_{\theta_1, \theta_2, r_A, r_I}[ f\otimes g(x)]\nonumber\\
&&\textnormal{subject to}\nonumber\\
&&r_A > \rho_A(\theta_1), r_I > \rho_I(\theta_2) \nonumber\\
&&r_A + r_I = c \nonumber\\
&&\theta_1, \theta_2>0.
\end{eqnarray}
In here, $\rho_A(\theta_1)$ ($\rho_I(\theta_2)$) is the function of
$\theta_1$ ($\theta_2$).

\subsection{Ciucu's Backlog Bounds}\label{subsec:ciucu_bound}

Because $A$ is $(\sigma_{A}(\theta_1), \rho_{A}(\theta_1))$-upper
constrained, by Theorem~\ref{theo:ac_comput}, $A \sim_{ta}
<f,\alpha>$, where
\begin{eqnarray}\label{eq:ciucu_A}
\alpha(t) &=& r_A \cdot t\nonumber\\
f(x) &=& e^{\theta_1\sigma_{A}(\theta_1)}\cdot e^{-\theta_1 x},
\end{eqnarray}
for any $r_A \geq \rho_{A}(\theta_1)$.

In the same way, $I \sim_{ta} <g,\gamma>$, where
\begin{eqnarray}\label{eq:ciucu_I}
\gamma(t) &=& r_I \cdot t\nonumber\\
g(x) &=& e^{\theta_2\sigma_{I}(\theta_2)}\cdot e^{-\theta_2 x},
\end{eqnarray}
for any $r_I \geq \rho_{I}(\theta_2)$.

By Lemma~\ref{lemma:ta_vb}, we have $A \sim_{vb}
<\tilde{f},\alpha_{\delta_1}>$ where
\begin{eqnarray}\label{eq:ciucu_A_vb}
\alpha_{\delta_1}(t) &=& (r_A+\delta_1) \cdot t\nonumber\\
\hat{f}(x,\delta_1) &=&
\frac{e^{\theta_1\sigma_{A}(\theta_1)}}{1-e^{-\theta_1 \delta_1}}
\cdot e^{-\theta_1 x},
\end{eqnarray}
for any $r_A \geq \rho_{A}(\theta_1)$ and $\delta_1>0$. In here, we
can get the close form of $\tilde{f}(x,\delta) =
\sum_{k=0}^{\infty}{f(x+k\delta})$ for the particular $f(x)$ in
Eq.(\ref{eq:ciucu_A}).

In the same way, $I \sim_{vb} <\tilde{g},\gamma_{\delta_2}>$ where
\begin{eqnarray}\label{eq:ciucu_I_vb}
\gamma_{\delta_2}(t) &=& (r_I+\delta_2) \cdot t\nonumber\\
\hat{g}(x,\delta_2) &=&
\frac{e^{\theta_2\sigma_{I}(\theta_2)}}{1-e^{-\theta_2 \delta_2}}
\cdot e^{-\theta_2 x},
\end{eqnarray}
for any $r_I \geq \rho_{I}(\theta_2)$ and $\delta_2>0$.

By Theorem~\ref{theo:ciucu_leftover}, the node provides a stochastic
service curve $S \sim <\tilde{g}, \beta>$, where
\begin{eqnarray}\label{eq:ciucu_S}
\beta_{-\delta_2}(t) &=& (c-r_I-\delta_2) \cdot t,
\end{eqnarray}
for any $c>r_I+\delta_2$.

Finally, by Theorem~\ref{theo:ciucu_backlog}, we must have
$\alpha_{\delta_1}(t) \leq \beta_{-\delta_2}(t)$, i.e., $r_A +
\delta_1 \leq c - r_I - \delta_2$ in order to get meaningful backlog
bounds which are $P\{B(t)>x\} \leq \tilde{f}\otimes \tilde{g}(x)$.
We note that $f(x)$ ($g(x)$) is the decreasing function of
$\delta_1$ ($\delta_2$). Considering the above conditions, we get
the following optimal backlog bounds,

\begin{eqnarray}\label{eq:ciucu_backlog}
&&P\{B(t) > x\} \leq \min_{\theta_1, \theta_2, \delta_1, \delta_2, r_A, r_I}[ \tilde{f}\otimes \tilde{g}(x)]\nonumber\\
&&\textnormal{subject to}\nonumber\\
&&r_A > \rho_A(\theta_1), r_I > \rho_I(\theta_2) \nonumber\\
&&r_A + r_I + \delta_1 + \delta_2= c \nonumber\\
&&\theta_1, \theta_2>0, \delta_1, \delta_2>0.
\end{eqnarray}

In here, $\rho_A(\theta_1)$ ($\rho_I(\theta_2)$) is the function of
$\theta_1$ ($\theta_2$).

\subsection{{\bf Equivalent Bounds in Two Models}}\label{subsec:equiv}

We find that the two models actually can derive the same stochastic
service curves and backlog bounds. To understand this, we note that
the key difference is the traffic model. The following proposition
shows that we can derive the same vb arrival curves from the two
models.

\begin{proposition}[ta vs vb Arrival Curves]\label{prop:2vb_equiv}
If $A$ is a $(\sigma(\theta), \rho(\theta))$-upper constrained
process, its vb arrival curve immediately generated by applying
Theorem~\ref{theo:ac_comput}'s Eq.(\ref{eq:vb_comput}) and the one
generated by applying Theorem~\ref{theo:ac_comput}'s
Eq.(\ref{eq:ta_comput}) and then Lemma~\ref{lemma:ta_vb} are
equivalent.
\end{proposition}
\noindent{\emph{Proof:}} Following the discussions above, the vb
arrival curves generated immediately by applying
Theorem~\ref{theo:ac_comput}'s Eq.(\ref{eq:vb_comput}) are
 $A \sim_{vb}<f,\alpha>$ where
\begin{eqnarray}\label{eq:prop_jiang_A}
\alpha(t) &=& r \cdot t\nonumber\\
f(x)&=& \frac{e^{\theta
\sigma(\theta)}}{1-e^{\theta(\rho(\theta)-r)}}\cdot e^{-\theta x},
\end{eqnarray}
for any $r > \rho(\theta)$.

The vb arrival curves by applying Theorem~\ref{theo:ac_comput}'s
Eq.(\ref{eq:ta_comput}) and then Lemma~\ref{lemma:ta_vb} (converted
by the ta arrival curves) are $A \sim_{vb}
<\tilde{f},\alpha_{\delta}>$ where
\begin{eqnarray}\label{eq:prop_ciucu_A_vb}
\alpha_{\delta}(t) &=& (r+\delta) \cdot t\nonumber\\
\hat{f}(x,\delta) &=& \frac{e^{\theta\sigma(\theta)}}{1-e^{-\theta
\delta}} \cdot e^{-\theta x},
\end{eqnarray}
for any $r \geq \rho(\theta)$ and $\delta>0$.

For the same value of $(r+\delta)$ in Eq.(\ref{eq:prop_ciucu_A_vb}),
we should maximize $\delta$ to get tighter $\hat{f}(x,\delta)$; in
other words, we should minimize $r$ and let it to be $\rho(\theta)$.
In this optimized case we find that Eq.(\ref{eq:prop_jiang_A}) and
Eq.(\ref{eq:prop_ciucu_A_vb}) are in the same form. This establishes
the equivalence between them. \done

This result can be applied to the vb arrival curves of impairment
processes. Since Ciucu's model can be derived from Jiang's for the
single-server case (Section~\ref{subsec:ciucu}), the two models can
derive the same backlog bounds in the following proposition.

\begin{proposition}[Bounds Equivalence in Two Models]\label{prop:model_equiv}
Consider a single server $S$ with an ideal service curve
$\hat{\beta}$ and an impairment process $I$. Suppose the traffic
arrival process $A$ and the impairment process $I$ are
$(\sigma(\theta), \rho(\theta))$-upper constrained for some $\theta$
respectively, then the vb arrival curves, the stochastic service
curve and backlog bounds derived by Jiang's and Ciucu's model are
equivalent.
\end{proposition}

We omit the proof which is got immediately from
Proposition~\ref{prop:2vb_equiv}.

\textbf{Note:} By equivalence, we do not mean it is general for all
situations. Actually, Ciucu's model can be extended to multiple
concatenated nodes while Jiang's can not, and they are different
models. Even for the single-node case, we only prove the equivalence
property for \emph{linear} arrival curves in our studied case. And
it is still an open problem for more general cases.

\subsection{{\bf Stability Condition}}\label{subsec:stability}

One fundamental question we need to address is under what condition
we can derive \emph{stable} backlog bounds (i.e., $\tE B(t) <
\infty$) by stochastic network calculus. The following proposition
shows the stability condition.

\begin{proposition}[Stability Condition]\label{prop:stability}
Suppose there exist $\theta$-MERs ($\theta$-Minimum Envelop Rates)
for the traffic arrival process $A$ and the impairment process $I$
of the wireless node for $0<\theta<\hat{\theta}$ where
$\hat{\theta}$ is some constant value, then stochastic network
calculus can derive stable backlogs if
\begin{eqnarray}\label{eq:stab}
a_{A} < c-a_{I},
\end{eqnarray}
where $c$ is the transmission rate of the ideal channel, $a_{A}$ and
$a_{I}$ are the average rate of $A$ and $I$ defined in
Definition~\ref{def:avg_rate}, respectively.
\end{proposition}

\noindent\emph{Proof:}

The proof consists of two phases. First, we show that $a_A < c -
a_I$ can lead to $r_A \leq c - r_I$. Next, we show that if $r_A \leq
c - r_I$ then stochastic network calculus can derive $\tE B(t)$
which is less than a finite value.

We adopt Jiang's model in Section~\ref{subsec:jiang_bound} where $A$
is the traffic arrival process and $I$ is the impairment process of
the server $S$, since the two models are equivalent in our studied
case (see Proposition~\ref{prop:model_equiv}). We have shown that
$P\{B(t)>x\} \leq f\otimes g(x)$ if $r_A \leq c - r_I$ holds.

From Eq.(\ref{eq:jiang_A}) and (\ref{eq:jiang_I}), we let
$\epsilon_1 = r_A - \rho_{A}(\theta_1)$ and $\epsilon_2 = r_I -
\rho_{I}(\theta_2)$ for $\theta_1, \theta_2>0$ and $\epsilon_1,
\epsilon_2>0$. To simplify the arguments, we let $\theta_1 =
\theta_2 = \theta$ and $\epsilon_1 = \epsilon_2 = \epsilon$.

Thus, $r_A \leq c - r_I$ holds if
\begin{eqnarray}\label{eq:stab_ER}
\rho_{A}(\theta) \leq c - \rho_{I}(\theta) - 2\epsilon.
\end{eqnarray}

From Theorem~\ref{theo:theta-ER}, we can construct the
$(\sigma(\theta), \rho(\theta))$-upper constrained characterization
by letting $\rho_{A}(\theta)=\rho_{A}^*(\theta)+\epsilon$ and
$\rho_{I}(\theta)=\rho_{I}^*(\theta)+\epsilon$ for any $\epsilon>
0$, where $\rho_{A}^*(\theta_1)$ and $\rho_{I}^*(\theta_2)$ are
$\theta$-MERs of $A$ and $I$, respectively. And
Eq.(\ref{eq:stab_ER}) holds if
\begin{eqnarray}\label{eq:stab_MER}
\rho_{A}^*(\theta) \leq c-\rho_{I}^*(\theta) - 4 \epsilon.
\end{eqnarray}

Because $\rho_{A}^*(\theta)$ exists, applying Taylor's expansion,
\begin{eqnarray*}
&&\rho_{A}^*(\theta) = \lim\sup_{t\rightarrow \infty}\frac{1}{\theta
t}\sup_{s\geq 0}\log{\tE e^{\theta A(s,s+t)}}\\
&&= \lim {\sup_{t\rightarrow \infty}{\frac{1}{\theta
t}\sup_{s\geq 0}{\log{\tE (1 + \theta A(s,s+t) + O(\theta^2 A(s,s+t)^2))}}}}\\
&&= \lim{\sup_{t\rightarrow \infty}{\frac{1}{\theta
t}\sup_{s\geq 0}{\log{(1 + \theta \tE A(s,s+t) + O(\theta^2 A(s,s+t)^2))}}}}\\
&&= \lim{\sup_{t\rightarrow \infty}{\frac{1}{\theta t}\sup_{s\geq
0}{[\theta \tE A(s,s+t) + O(\theta^2 A(s,s+t)^2)]}}}.
\end{eqnarray*}

Let $\theta$ go to 0,
\begin{eqnarray}
\lim_{\theta \rightarrow 0}\rho_{A}^*(\theta) = \lim_{t\rightarrow
\infty}\sup_{s\geq 0}{\frac{\tE A(s,s+t)}{t}} = a_{A}.
\end{eqnarray}

Similarly,
\begin{eqnarray}
\lim_{\theta \rightarrow 0}\rho_{I}^*(\theta) = a_{I}.
\end{eqnarray}

Therefore, there exists some $\theta < \hat{\theta}$ so that
$\rho_{A}^*(\theta)\leq a_{A}+\epsilon$ and $\rho_{I}^*(\theta)\leq
a_{I}+\epsilon$. So Eq.~(\ref{eq:stab_MER}) holds if
\begin{eqnarray}\label{eq:stab_a1}
a_{A} \leq c-a_{I} - 6\epsilon.
\end{eqnarray}

Since $\epsilon$ can be arbitrarily small, Eq.(\ref{eq:stab_a1})
holds if
\begin{eqnarray}\label{eq:stab_a}
a_{A} < c-a_{I}.
\end{eqnarray}

Following the above derivations backwards, we prove that $a_{A} <
c-a_{I}$ leads to $r_{A} < c-r_{I}$.

Next, we prove that stochastic network calculus can derive $\tE
B(t)$ which is less than a finite value if $r_A < c - r_I$.

Since $f(x)$ and $g(x)$ are exponentially decreasing functions
according to Eq.~(\ref{eq:jiang_A}) and Eq.~(\ref{eq:jiang_I}), we
can show that $\tE B(t)$ is upper-bounded by some finite constant
value as follows. Note that $B(t)$ is a discrete value in practice
(e.g., in bits or packets).

\begin{eqnarray}
&&\tE B(t) = \sum_{k=0}^\infty P\{B(t) = k+1\} \cdot (k+1)\nonumber\\
&&< \sum_{k=0}^\infty P\{B(t) > k \} \cdot (k+1)\nonumber\\
&&\leq \sum_{k=0}^\infty f\otimes g(k) \cdot (k+ 1) \nonumber\\
&&\leq \sum_{k=0}^\infty (f(\lfloor\frac{k}{2}\rfloor) +
g(\lceil\frac{k}{2}\rceil)) \cdot (k + 1) < \infty.
\end{eqnarray}

\done
\\

\noindent\textbf{Remarks:} Since the proof is based on the theory of
stochastic network calculus, it indicates that we can get stable
backlog bounds by stochastic network calculus on the condition that
the average arrival rate is less than the average service rate. As
this condition is very general, stochastic network calculus is
effective in theory.

\section{{\bf An 802.11 Node's Network Calculus Model}}\label{sec:802_11}

In this section, we derive the backlog bounds for an 802.11 node.
And the key part is to derive its stochastic service curve. We use
Jiang's model in this section since the two models are equivalent in
our studied case (Proposition~\ref{prop:model_equiv}).

For simplicity, we assume $n$ \emph{identical} nodes send packets to
an AP (access point) and they share the wireless channel. All nodes
operate in Distributed Coordination Function (DCF) mode with RTS/CTS
turned off\cite{802_11}. We assume that transmission errors only
happen due to packet collisions and two packets are collided if
their transmissions overlap in time. Besides, we assume that all
DATA packets are of the same size for simplicity. We use Scenario 1
for illustration.

\begin{figure}[htb]
\begin{center}
\begin{tabular}{|l|l|}
\hline
\textbf{Scenario 1:}\\
\ 10 nodes send packets to one AP in a WLAN\\
\ The payload of a DATA packet is $256$ bytes\\
\hline
\end{tabular}
\end{center}
\caption{Scenario 1 of a wireless LAN}\label{fig:scen1}
\end{figure}

\subsection{{\bf 802.11 DCF Protocol}}\label{subsec:802_11_DCF}
A node with a DATA packet (or simply packet) to transmit first
senses channel state. If the channel is idle for the time of DIFS
(distributed interframe space), the node transmits. Otherwise, if
the channel is busy during the DIFS, the node backs off, i.e., the
node defers channel access by a random number of \emph{idle slots}
ranging from 0 to $CW-1$ within a contention window ($CW$). When the
backoff counter reaches zero and expires, the node can access the
channel. During the backoff period, if the node senses the channel
is busy, it freezes the backoff counter and the backoff process is
resumed once the channel is idle for a duration of DIFS. To avoid
channel capture, a node must wait a random backoff time between two
consecutive new packet transmissions, even if the channel is sensed
idle. Once the packet is received successfully, the receiver will
return an ACK after the duration of SIFS (short interframe space).
SIFS is shorter than an idle slot so that there are no collisions
caused by DATA packets and ACKs.

802.11 uses the truncated exponential backoff technique to set its
$CW$. In 802.11b, the initial $CW$ is $CW_{min}=32$. Each time a
collision occurs, $CW$ doubles its size, up to $CW_{max}=1024$. When
the packet is successfully transmitted, $CW$ is reset to $CW_{min}$.
The packet is dropped when it is retransmitted for 6 times and still
not transmitted successfully. Fig.~\ref{fig:802_11} shows the
parameters of 802.11b used in our paper.

\begin{figure}[htb]
\begin{center}
\begin{tabular}{|l|l|}
\hline
Basic rate & 1 Mbps\\
Data rate & 11 Mbps\\
PHY header & 24 bytes\\
ACK header & 14 bytes\\
MAC header & 28 bytes\\
SIFS & 10 $\mu$s\\
DIFS & 50 $\mu$s\\
Idle slot & 20 $\mu$s\\
$CW_{min}$ & 32\\
$CW_{max}$ & 1024\\
\hline
\end{tabular}
\end{center}
\caption{802.11b parameters}\label{fig:802_11}
\end{figure}

The duration of an ACK is the duration of PHY header plus that of
ACK header transmitted at \emph{basic rate}, i.e.,
$\frac{(24+14)\cdot 8}{10^6} = 304\mu s  \approx 16 \ idle\ slots$.
The duration of a DATA packet is the duration of PHY header
transmitted at \emph{basic rate} plus that of an MAC header and its
upper-layer payload transmitted at \emph{data rate}. For example,
suppose the upper-layer payload is $256$ bytes, then the duration of
an DATA packet is $\frac{24\cdot 8}{10^6}+\frac{(28+256) \cdot
8}{11\cdot 10^6}=398.5\mu s \approx 20 \ idle\ slots$.

\subsection{{\bf 802.11 Service Curve}}\label{subsec:802_11_sc}
Since equal-sized DATA packets are considered, \emph{we measure
traffic, service and backlog amount in packets in our paper.} We
measure time duration (e.g. SIFS, DIFS, DATA and ACK) in the unit of
idle slots and define that \emph{one time slot of network calculus
($t=1$) is equal to $L$ idle slots}, where
\begin{eqnarray}\label{eq:L}
L = (DIFS+DATA+SIFS+ACK) \ in \ idle\ slots.
\end{eqnarray}
Ideally, an 802.11 node transmits 1 packets per time slot ($L$ idle
slots in length). Suppose the DATA payload is $256$ bytes,
$L=3+16+20=39\ idle\ slots$.

Sometimes in the paper, \emph{"idle slot" refers to the time period
which equal to the length of an idle slot and it may not be idle.}
To avoid this confusion, we will use "\emph{idle slot}" (italic) to
denote that the "idle slot" is indeed idle.

An 802.11 node can be modeled as an ideal server (1 packet
transmitted per time slot) with the impairment process $I$ which is
due to contention with the other nodes in a WLAN. In practice, it is
difficult to calculate $I$ accurately since $I$ depends on the
complex interactions of traffic arrival and DCF. In this section, we
assume the saturated state and use A. Kumar's fixed-point model of
802.11\cite{kumar_802_11}. This model is justified to be very
accurate in practice \cite{gaoyan_802_11}.

We assume that the system is working at the saturated state, that
is, the backlog at each node is always nonzero. For a given node,
let $\tau$ denote its transmission attempt probability per
\emph{idle slot} and let $\eta$ denote the conditional collision
probability when it transmits a packet. We assume $\eta$ is constant
and independent for each transmission. Intuitively, this assumption
becomes more accurate when the number of nodes $n$ increases. In
\cite{kumar_802_11}, the authors derived two general formulas
relating $\tau$ to $\eta$. The first one is

\begin{eqnarray}\label{eq:tau_eta}
\tau = \frac{1+\eta+\eta^2+...+\eta^6}{b_0+\eta b_1+\eta^2
b_2+...+\eta^6 b_6}.
\end{eqnarray}

This equation can be explained as follows. The numerator is the
expected number of transmission attempts of a packet. The
denominator is the expected total backoff duration (in idle slots)
of a packet, where $b_i$ is the mean backoff duration after the
$i$th collision plus 1 (the 1 refers to the first idle slot of a
packet transmission). In 802.11, $b_i = \frac{2^i \cdot
CW_{min}}{2}$ where $0 \leq i \leq 6$. A packet suffering 6
consecutive collisions will be dropped from its buffer. In our
calculations, we do not consider packet drops. Since the probability
of packet drops is very small, this simplification relaxes the
backlog bounds very slightly.

The independence assumption of $\eta$ implies that each transmission
sees the system at steady state. Therefore, each node transmits with
the same probability $\tau$. This yields
\begin{eqnarray}\label{eq:eta_tau}
\eta = 1-(1-\tau)^{n-1}.
\end{eqnarray}
Combining Eq.~(\ref{eq:tau_eta}) and Eq.~(\ref{eq:eta_tau}), we can
solve $\tau$ and $\eta$.

We introduce the following terms. The probability of no
transmissions at an \emph{idle slot} in the WLAN, denoted by
$P_{nt}$, is $(1-\tau)^n$. The probability of having at least one
transmission at an \emph{idle slot} in the WLAN, denoted by $P_t$,
is $1-P_{nt}$. The probability of \emph{a given node} starting a
successful transmission at an \emph{idle slot}, denoted by $P_s$, is
$\tau (1-\eta)$.

\begin{figure}[htb]
  \centering
  \includegraphics[width=0.8\textwidth]{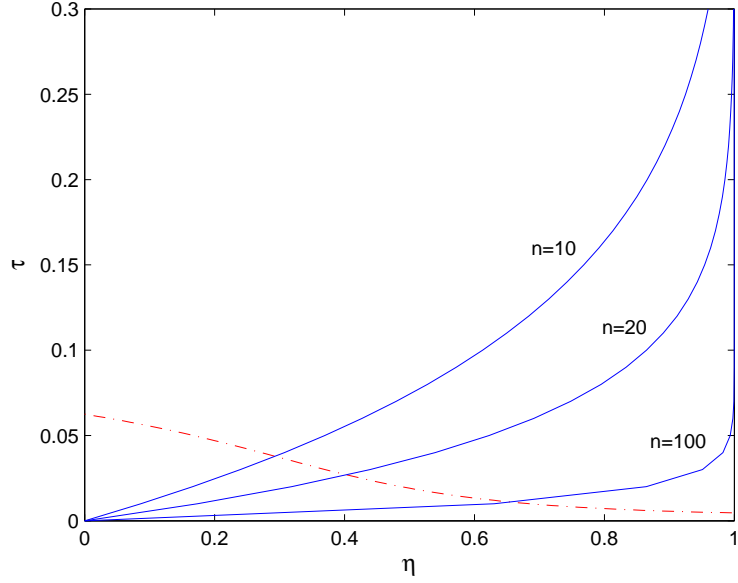}\\
  \caption{The Plots of Eq.~(\ref{eq:tau_eta}) and Eq.~(\ref{eq:eta_tau})}\label{fig:eta_tau}
\end{figure}

Fig.~\ref{fig:eta_tau} plots Eq.~(\ref{eq:tau_eta}) in dashed line
and Eq.~(\ref{eq:eta_tau}) in solid line when $n = 10$, $20$ and
$100$. The intersecting points are the solutions to $\eta$ and
$\tau$. It can be seen from the figure that $\eta$ increases and
$\tau$ decreases as $n$ increases. Consequently, $P_{nt}$ increases
while $P_s$ and $P_t$ decreases as $n$ increases. When we consider
the saturated state of the system, we actually consider all nodes
contending the channel which gives the \emph{worst-case} analysis of
the impairment process of a given node and thus conservative backlog
bounds of it. However, we argument that it is necessary because one
applies network calculus to deriving the worst-case bounds.

In order to characterize the impairment process $I$ of the given
802.11 node, it is crucial to know its moment generating function.
Specifically, we want to calculate
\begin{eqnarray}\label{eq:M_I}
M_I(t)=\sup_{s\geq 0}[\tE e^{\theta I(s,s+t)}],
\end{eqnarray}
and then we can know its $(\sigma(\theta), \rho(\theta))$-upper
constrained characterization (we replace $\theta_2$ by $\theta$ here
to simplify explanations).

We calculate Eq.(\ref{eq:M_I}) for a given 802.11 node as follows.
Consider the duration of $t$ time slots (i.e., $tL$ idle slots) from
$s$ to $s+t$. Since we take the sup, we can assume that there is
always a transmission by the other nodes at the first time slot
$[s,s+1]$, which gives the conservative estimation of $I(s,s+t)$ for
the given node. We can see that it is a good approximation as the
transmissions by the other $n-1$ nodes happen much frequently than
the given node when $n$ is large and also an \emph{idle slot} is
much smaller in length than $L$ (in other words, the channel is
often busy). We can see this point is right for Scenario 1 in the
end of this subsection.

In the following, we consider probabilistic events in the remaining
$t-1$ time slots. There are two cases. \emph{Case I:} The last
transmission is "incomplete". \emph{Case II:} Otherwise to Case I.
By "incomplete", we means that the last transmission goes on for k
idle slots ($1\leq k \leq L-1$) and get truncated due to the
boundary of the last time slot. Let $\tilde{P}_s = P_s/P_t$ denote
the condition probability of the given node's successful
transmission on the condition that there is a transmission on the
channel.

We first calculate $M_I(t)$ for case I. Suppose there are $i$
complete transmissions and one incomplete transmission occupying k
idle slots, its probability denoted by $p_{i,k}$ is $P_t \cdot
C_{(t-i-1)L-k+i}^{i}P_t^i P_{nt}^{(t-i-1)L-k}$. In here, we use the
fact that there are $i$ complete transmissions, 1 incomplete
transmission and thus $(t-i-1)L-k$ idle slots in the remaining $t-1$
time slots. Suppose there are $j$ successful transmissions from the
given node in the $i$ complete transmissions, its probability
denoted by $q_{j,i}$ is $C_i^j (\tilde{P}_s)^j
(1-\tilde{P}_s)^{i-j}$. When the last incomplete transmission is
from the other nodes, $M_I(t)$ is $e^{\theta(t-j)}$; otherwise, the
transmission is from the given node itself, $M_I(t)$ is
$e^{\theta(t-j-k/L)}$.

Numerating all possible $k$, $i$ and $j$, $M_I(t)$ under case I is
{\setlength\arraycolsep{2pt}\begin{eqnarray}\label{eq:M_I1}
&&\sum_{k=1}^{L-1}\sum_{i=0}^{t-2}\sum_{j=0}^{i} \big( p_{i,k}
q_{j,i}\tilde{P}_s e^{\theta(t-j-k/L)} + p_{i,k}
q_{j,i}(1-\tilde{P}_s) e^{\theta(t-j)} \big) \nonumber\\
&&=\sum_{k=1}^{L-1}\sum_{i=0}^{t-2} p_{i,k}  (\tilde{P}_s
e^{-\theta k/L}+1-\tilde{P}_s)  (\tilde{P}_s e^{-\theta}+1-\tilde{P}_s)^i  e^{\theta t}\nonumber\\
\end{eqnarray}}

Then we calculate $M_I(t)$ for case II. Let
$p_i=C_{(t-i-1)L+i}^{i}P_t^i P_{nt}^{(t-i-1)L}$. Following the
similar arguments as above, $M_I(t)$ for case II is:
\begin{eqnarray}\label{eq:M_I2}
\sum_{i=0}^{t-1} \sum_{j=0}^{i} p_i q_{j,i} e^{\theta
(t-j)}=\sum_{i=0}^{t-1} p_i (\tilde{P}_s
e^{-\theta}+1-\tilde{P}_s)^i e^{\theta t}.
\end{eqnarray}
Adding Eq.(\ref{eq:M_I1}) and Eq.(\ref{eq:M_I2}), finally we get
$M_I(t)$.

In general, we do not have the analytical form of $M_I(t)$, so we
resort to numerical methods to obtain $\sigma_{I}(\theta)$ and
$\rho_{I}(\theta)$ (see Algorithm 1 in Appendix B). The algorithm is
immediately inspired from Definition~\ref{def:sigma_rho}. Then we
can use Eq.~(\ref{eq:jiang_I}) and (\ref{eq:jiang_S}) to obtain the
node's stochastic service curve.

We illustrate the above calculations for Scenario 1 in
Fig.~\ref{fig:scen1}. From Eq.~(\ref{eq:tau_eta}) and
(\ref{eq:eta_tau}), $\tau = 0.037$ and $\eta = 0.293$. Thus,
$P_{nt}=0.680$, $P_{t}=0.320$ and $P_s = 0.027$. Again, we can see
that the previous assumption that the first slot in $[s,s+t]$ is
occupied by a transmission from the other $n-1$ node is a good
approximation, as the transmissions by the other $n-1$ nodes happen
much frequently than the given node when $n$ is large (compare
$P_t-P_s$ and $P_s$ here) and also an \emph{idle slot} is much
smaller in length than $L$ ($39$ idle slots here; in other words,
the channel is often busy).

\begin{figure}[htb]
  \centering
  \includegraphics[width=0.8\textwidth]{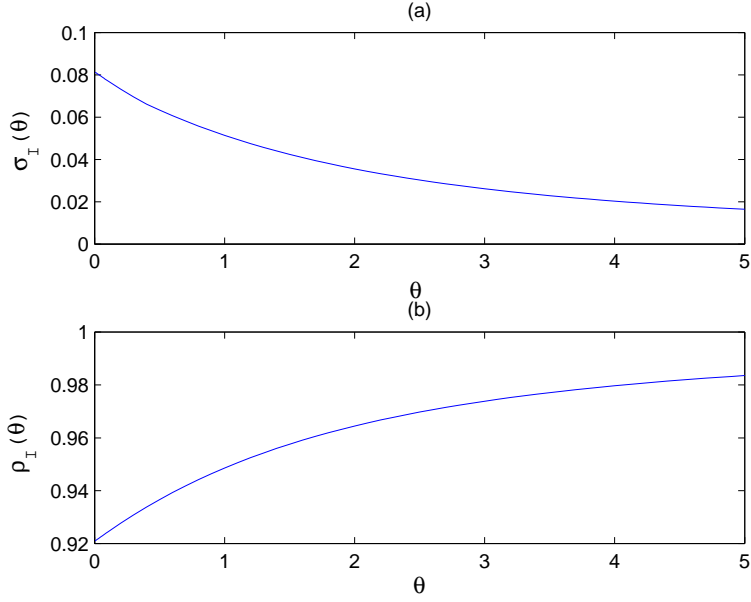}\\
  \caption{The impairment process $I$'s $(\sigma(\theta), \rho(\theta))$-upper constrained characterization}\label{fig:I_up_cons}
\end{figure}

Fig.~\ref{fig:I_up_cons} shows  $I$'s $\sigma(\theta),
(\rho(\theta))$-upper constrained characterization when $\theta$
ranges from 0.01 to 5.0.

For example, when $\theta=0.1$, we have $\sigma_I(0.1)=0.077$ and
$\rho_I(0.1)=0.924$. And
\begin{eqnarray*}
\beta(t) &=& (1-r_I)\cdot t\\
g(x) &=& \frac{e^{0.0077}}{1-e^{0.0924-0.1r_I}}\cdot e^{-0.1 x},
\end{eqnarray*}
for any $1 > r_I > 0.924$.

Finally, we can not apply Proposition~\ref{prop:vb_improve} to
tightening $g(x)$, as $\rho(\theta)+\sigma(\theta) \geq 1$ here
(considering $M_I(1)=e^\theta \leq
e^{\rho(\theta)+\sigma(\theta)}$). In order to apply this
proposition, we must let $r_I \geq \rho(\theta)+\sigma(\theta) \geq
1$ which makes $\beta(t)=(1-r_I)t \leq 0$.

\subsection{{\bf Arrival Curves}}\label{subsec:802_11_ac}
In our performance evaluation we use Poisson traffic and we let
$\lambda$ be the average rate (packets/slot) of it. We have $a_A =
\lambda$ by Definition~\ref{def:avg_rate}, and
\begin{eqnarray}
Ee^{\theta A(s,s+t)} = \sum_{i=0}^\infty \frac{(\lambda t)^i}{i!}
e^{-\lambda t} e^{\theta i} = e^{\lambda t(e^\theta-1)}.
\end{eqnarray}. Therefore, Poisson traffic is $(\sigma_A(\theta),
\rho_A(\theta)$-upper constrained where $\sigma_A(\theta)=0$ and
$\rho_A=\frac{\lambda(e^\theta-1)}{\theta}$.

We can get the vb arrival curves by Eq.~(\ref{eq:jiang_A}). Note
that we can improve the bounding function by $f(x)=e^{-\theta x}$
for $r_A \geq \rho_A(\theta)$ by Proposition~\ref{prop:vb_improve}
as Poisson process is time-independent.

As for the traffic in reality, we can get $\tE e^{\theta A(s,s+t)}$
from traffic traces. Then we use Algorithm 1 to get the
$(\sigma_A(\theta), \rho_A(\theta)$-upper constrained
characterization and the vb arrival curves.

\subsection{{\bf Stability Condition and Backlog Bounds}}\label{subsec:802_11_backlog}

By Proposition~\ref{prop:stability}. the \emph{stability condition
of an 802.11 node in a WLAN} is
\begin{eqnarray}\label{eq:stab_802_11}
a_A < 1-a_I = \frac{P_s \cdot L}{P_{nt} + P_t \cdot L}.
\end{eqnarray}

The stability condition in Scenario 1 is $a_A < 0.079$ packet/slot
or 0.207Mbps by the 802.11 parameters in Fig.~\ref{fig:802_11}.

The backlog bounds is calculated immediately by
Eq.(\ref{eq:jiang_backlog}) by plugging into traffic arrival curves
and the 802.11 node's service curve. Note that it is an optimization
problem depending on $\theta_1$ and $\theta_2$. In general, we do
not have an analytical solution for it. Since the problem dimension
is very small, we can apply the method of exhaustion to get the
optimal value.

We can also use Theorem~\ref{theo:independent} to improve on backlog
bounds, as in our model we consider $I$ under the saturated state
which is independent of $A$. And it still needs to optimize the
derived bounding function.

\section{{\bf Performance Evaluation}}\label{sec:simulation}

In this section, we compare our backlog bounds derived in
Section~\ref{sec:802_11} with ns-2 simulations in Scenario 1 with
Poisson traffic arrivals. The duration of each ns-2 simulation is
100 seconds which is long enough to let a node transmit thousands of
packets. And we get the real $P\{B(t)>x\}$) over 100 independent
simulations.

\begin{figure}[htb]
  \centering
  \includegraphics[width=0.8\textwidth]{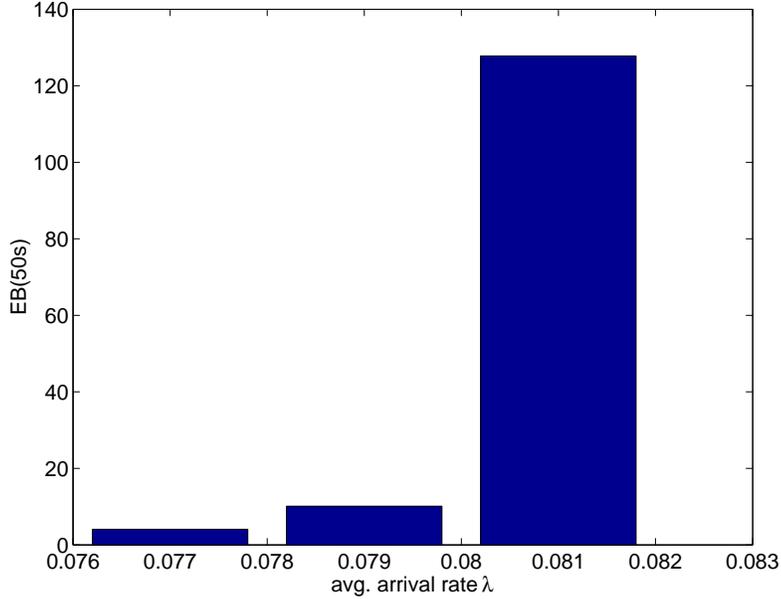}\\
  \caption{$\tE B(t)$ ($t=50$s) when $\lambda=0.077, 0.079, 0.081$}\label{fig:plot_10_stable_Poisson}
\end{figure}

As shown in Section~\ref{subsec:802_11_backlog}, we can derive
stable backlog bounds when $a_A=\lambda < 0.079$ packet per slot.
Fig.~\ref{fig:plot_10_stable_Poisson} plots the average backlog
$\tE[B(t)]$ of ns-2 at $t=50s$ and $\lambda = 0.077$, $0.079$ and
$0.081$ packet/slot. We note that there is a sudden jump when
$\lambda = 0.081$, indicating the critical point of stability is
indeed around $0.079$.

We use Jiang's model to calculate backlog bounds since the two
models are proved to be equivalent in our studied case. There are
two results of vb arrival curves for the traffic arrival process
$A$: the \emph{general} one (Theorem~\ref{theo:ac_comput}'s
Eq.(\ref{eq:vb_comput})) and the \emph{time-independent} one which
improves on the former for the time-independent $A$
(Proposition~\ref{prop:vb_improve}). And there are two relations
between $A$ and the impairment process $I$ of the server: the
\emph{general} one (Theorem~\ref{theo:jiang_backlog}) and the
\emph{independent} one which improves on backlog bounds for
independent $A$ and $I$ (Theorem~\ref{theo:independent}).

So we can generate four results of backlog bounds.
\begin{itemize}
\item Bound 1: by \emph{general} vb arrival curves of traffic and \emph{general} backlog bounds.
\item Bound 2: by \emph{time-independent} vb arrival curves of traffic and \emph{general} backlog bounds.
\item Bound 3: by \emph{general} vb arrival curves of traffic and \emph{$A$-$I$-independent} backlog bounds.
\item Bound 4: by \emph{time-independent} vb arrival curves of traffic and \emph{$A$-$I$-independent} backlog bounds.
\end{itemize}
Obviously, Bound 1 is the loosest and Bound 4 is the tightest among
them.

To illustrate our results, we show the smallest $x$ that makes
$P\{B(t)>x\}\leq p$ for some probability $p$, i.e.,
$\min\{x:P\{B(t)>x\}\leq p\}$ where $p=0.9,0.8,...,0.1,0.05$.
Obviously, for the same $p$, smaller $x$, tighter the bound.
Fig.~\ref{fig:sim_low} (Fig.~\ref{fig:sim_high}) shows our results
when the 802.11 node's traffic arrival rate is $\lambda=0.04$
($0.07$) packet/slot.

\begin{figure}[htb]
\begin{center}
\begin{tabular}{|l|l|l|l|l|l|}
\hline
\ & \multicolumn{5}{|c|}{$\min\{x:P\{B(t)>x\}\leq p\}$}\\
\cline{1-6}
$p$ & Bound 1 & Bound 2 & Bound 3 & Bound 4 & ns-2\\
\hline
0.9&24&8&7&7&1\\
\hline
0.8&25&9&8&8&1\\
\hline
0.7&25&10&8&8&1\\
\hline
0.6&25&10&9&9&1\\
\hline
0.5&26&11&10&10&1\\
\hline
0.4&27&12&10&10&1\\
\hline
0.3&28&13&11&11&1\\
\hline
0.2&29&14&12&12&2\\
\hline
0.1&31&17&14&14&2\\
\hline
0.05&33&19&16&16&4\\
\hline
\end{tabular}
\end{center}
\caption{ns-2 and network calculus results of backlog bounds under
$\lambda=0.04$packet/slot (low traffic load)}\label{fig:sim_low}
\end{figure}

%
%
%
%

\begin{figure}[htb]
\begin{center}
\begin{tabular}{|l|l|l|l|l|l|}
\hline
\ & \multicolumn{5}{|c|}{$\min \{x:P\{B(t)>x\}\leq p\}$}\\
\cline{1-6}
$p$ & Bound 1 & Bound 2 & Bound 3 & Bound 4 & ns-2\\
\hline
0.9&201&61&50&50&1\\
\hline
0.8&203&64&55&55&1\\
\hline
0.7&206&68&58&58&1\\
\hline
0.6&209&72&63&63&2\\
\hline
0.5&212&76&66&66&2\\
\hline
0.4&217&82&71&71&2\\
\hline
0.3&223&89&77&77&4\\
\hline
0.2&231&99&85&85&4\\
\hline
0.1&245&114&98&98&6\\
\hline
0.05&258&129&109&109&7\\
\hline
\end{tabular}
\end{center}
\caption{ns-2 and network calculus results of backlog bounds under
$\lambda=0.07$packet/slot (severe traffic load)}\label{fig:sim_high}
\end{figure}

%
%
%
%
%

We make the following observations. First, the backlog bounds
improve significantly when we apply \emph{time-independent} vb
arrival curves for traffic or $A$-$I$ independent case analysis.
Second, the bounds of network calculus are much looser for higher
traffic arrival rate while the real bounds of ns-2 simulations do
not relax much. Note that A. Kumar's 802.11 model become very
accurate near the saturated state \cite{gaoyan_802_11} which is just
the case here. The actual reason is: We have the constraint of
$\rho_A(\theta_1) + \rho_I(\theta_2) < c$
(Eq.(\ref{eq:jiang_backlog})). And we need to make $\theta_1$ and
$\theta_2$ smaller to satisfy this constraint for higher traffic
arrival rate, which leads to much looser bounding functions.
Moreover, Theorem~\ref{theo:ac_comput}'s Eq.(\ref{eq:vb_comput})
applies Boole's inequality to the bound functions of $I$, which are
loose in general. Here brings the challenge for better network
calculus models. Finally, we found in trace files that backlog
bounds are sensitive to the parameters (i.e., $\theta_1$,
$\theta_2$, $r_A$ and $r_I$) and it is necessary to optimize them.

\section{{\bf Related Work}}\label{sec:related}
In this section, we first present a brief overview of the theories
and applications of stochastic network calculus and then the related
works on the performance analysis of 802.11.

The increasing demand on transmitting multimedia and other real time
applications over the Internet has motivated the study of quality of
service guarantees. Towards it, deterministic and stochastic network
calculus has been recognized by researchers as a promising step.

Essentially, the network calculus is the theory of queueing systems
that comes from the seminal work by Cruz on the $(\sigma, \rho)$
traffic characterization \cite{netcal_cruz_1} \cite{netcal_cruz_2}
and work on the service curve characterization of Generalized
Processor Sharing (GPS) schedulers\cite{gps1}\cite{gps2}. The theory
has been developed by many researchers since then. The elegance of
network calculus is due to the fundamental convolution formulas
(under the min-plus algebra) that determine the departure process of
a system from its arrivals and its service curve. The notable
strength of the min-plus convolution is the ability to concatenate
tandem nodes along a network path, and therefore network calculus
has the ability to characterize the whole network as a single
server, which is generally intactable by traditional queueing theory
\cite{queueing_book}. Le Boudec's book covers deterministic network
calculus and its applications in the Internet\cite{boudec_book}.
Chang's book substantially presented the first approaches to
stochastic network calculus besides deterministic network
calculus\cite{chang_book}. Jiang summarized different types of
stochastic arrival and service curves in a unified framework and
proposed a new stochastic network calculus model stemmed from mb
(maximal backlog centric) arrival curves, although its application
conditions have some unsolved controversy. Jiang also wrote a book
on the theory of stochastic network calculus\cite{jiang_book}. Ciucu
proposed an effective stochastic service curve that can be applied
to concatenated systems and calculating end-to-end delay and backlog
bounds, which exhibits a good scaling property of $O(H\log H)$ where
$H$ is the number of nodes traversed by a flow\cite{ciucu_snc}.
Ciucu also showed that his model can derive quite accurate delay
bounds in M/M/1 and M/D/1 queueing systems by using the martingale
technique\cite{ciucu_martingale}. More recently, Fidler proposed a
novel solution of the queue system using expectations instead of
probabilities\cite{app_80211_1}, and he also made a comprehensive
survey on the recent progress of stochastic network calculus
\cite{fidler_overview}. Besides, Jiang wrote an overview on this
topic from the queueing principle perspective and he presented a
nice outlook by discussing many open
challenges\cite{jiang_overview}.

Many works have applied network calculus, for example, in
measurement-based admission control schemes \cite{app_mbac}, in
conformance testing, \cite{app_comformance}, in wireless sensor
networks\cite{app_sensor1}, in Aloha systems\cite{app_nonasymp}, in
speeding up network simulations\cite{app_tcpsim1}\cite{app_tcpsim2},
in bandwidth estimation\cite{app_bandwidth} and even in
manufacturing blocking systems in management science\cite{app_mbs}.

Compared with the existing theories of stochastic network calculus,
we study the effectiveness of backlog bounds in a practical 802.11
WLAN by using the two classic stochastic network calculus models:
Jiang's and Ciucu's. The latter can be extended to concatenated
systems while the former still can not at the moment. Interestingly,
we find that the two models can derive equivalent stochastic service
curves and backlog bounds in our studied case, which can provide
some hints for unifying the theories of stochastic network calculus
in the future.

Existing works on the performance of 802.11 focus primarily on the
throughput and capacity. Bianchi proposed a Markov chain model of
802.11\cite{bianchi_802_11}. A. Kumar et al. proposed a probability
model of 802.11\cite{kumar_802_11} which simplifies Bianchi's model
and it is shown to be quite accurate even in the multi-hop
case\cite{gaoyan_802_11}. In our paper, we adopt A. Kumar's model to
derive the stochastic service curves of an 802.11 node. There are
some works on 802.11 queueing analysis based on traditional queueing
theory. Zhai et al. assumed Poisson traffic arrival and proposed an
M/G/1 queueing model of 802.11\cite{zhai_802_11}. Tickoo proposed a
G/G/1 queueing model of
802.11\cite{tickoo1_802_11}\cite{tickoo2_802_11}. Bredel and Fidler
modeled the 802.11 DCF as a fluid GPS scheduler yielding a fair
average service rate \cite{app_80211_1}. And Ciucu analyzed the
non-asymptotic throughput and delay distribution in multi-hop
wireless networks by network calculus approach considering Aloha
systems\cite{app_nonasymp}.

Compared to existing analysis of 802.11, we are the first to analyze
the \emph{concrete} 802.11 transmissions by Jiang's and Ciucu's
models and study their \emph{effectiveness} on bounding backlogs. We
show that stochastic network calculus is effective
\emph{theoretically} in that the bounds imply stable backlogs as
long as the average arrival rate is less than the average service
rate. However, the bounds are quite loose and we show that they can
be improved significantly for time-independent arrivals or under the
independent cases of arrival and service processes. And we note that
it is still a challenge for a better theory of stochastic network
calculus towards tighter bounds in practice. Therefore, our work
offers a good reference to applying stochastic network calculus to
practical scenarios.

\section{{\bf Conclusion and Future Work}}\label{sec:conclusion}
In this paper, we present concrete computations of 802.11 backlog
bounds and study the bounds effectiveness using stochastic network
calculus, from general models to detailed calculations. We model a
wireless node as a single server with impairment service based on
two best-known models in stochastic network calculus:
Jiang's\cite{jiang_snc3} and Ciucu's\cite{ciucu_snc}. And we find
that they can derive equivalent stochastic service curves and
backlog bounds in our studied case.
Then we care about the effectiveness of network-calculus backlog
bounds \emph{theoretically}. And we prove that the network-calculus
backlog bounds imply stable backlog as long as the average rate of
traffic arrival is less than that of service.
Next, we consider the effectiveness of network-calculus bounds in
practice. We derive the stochastic service curve of an 802.11 node
from A. Kumar's 802.11 model, which is crucial to get backlog
bounds.
We observe the derived bounds are loose when compared with ns-2
simulations. However, the martingale and independent case analysis
techniques can improve the bounds significantly. But still the
bounds are not tight. We note the reason is due to the looseness in
network calculus itself such as Theorem~\ref{theo:ac_comput}'s
Eq.(\ref{eq:vb_comput}). The open questions are: Can we find tighter
bounding functions under certain conditions? How do we optimize on
stochastic arrival/service curves? Furthermore, does there exist any
unified theory of all network calculus models? And these are the
future works.

\section{Acknowledgement}
We thank Prof. Yuming Jiang with NTNU, Prof. Florin Ciucu with
TU-Berlin, and Dr. Kai Wang with Tsinghua and Caltech for their nice
discussions on the theories of stochastic network calculus. We thank
Prof. Jianming Zhu and Prof. Haiqi Feng in our school for holding
related seminars.


\begin{center}
    {\bf Appendix A: Proof of Proposition~\ref{prop:vb_improve}}
\end{center}

\emph{Proof:}

For a fixed t, we construct a stochastic process
$X(s)=e^{\theta(A(t-s,t)-rs)}$ ($0 \leq s\leq t$) and we have
$X(s+1)=X(s) e^{\theta(A(t-s-1,t-s)-r)}$. We will show that if
$a(s)\equiv A(s)-A(s-1)$ is independent for each time slot $s$ and
$r \geq \rho(\theta)+\sigma(\theta)$, then $X(s)$ is
supermartingale, i.e., $\tE [X(s+1)|X(0),...,X(s)] \leq X(s)$.

Because $a(s)$ is time-independent, we have
\begin{eqnarray}
&&\tE [X(s+1)|X(0),...,X(s)]= X(s) \cdot \tE [ e^{\theta(a(t-s)-r)}] \nonumber \\
&&=X(s) \cdot e^{-\theta r} \cdot \tE e^{\theta
a(t-s)}.\label{eq:iidE}
\end{eqnarray}

Because $A(t)$ is $(\sigma(\theta),\rho(\theta))$-upper constrained,
we have $\tE e^{\theta a(s)} \leq e^{\rho(\theta) + \sigma(\theta)}$
for all $s \geq 0$. When $r \geq \rho(\theta)+\sigma(\theta)$ and by
Eq.~(\ref{eq:iidE}), we have
\begin{eqnarray}
\tE [X(s+1)|X(0),...,X(s)] \leq X(s).
\end{eqnarray}
Thus, $X(s)$ is a supermartingale.

Doob's martingale inequality says that $P\{\sup_{0\leq s\leq t} X(s)
\geq k\} \leq \frac{\tE X(0)}{k}$ when $X(s)$ is a supermartingale
(note: $\tE X(0)=1$ here) for any constant $k$. Let $k=e^x$, we have
\begin{eqnarray}
&&P\{\sup_{0\leq s\leq t}[(A(s,t)-r(t-s)] > x\} \nonumber\\
&&=P\{\sup_{0\leq s\leq t} [X(s)] \geq e^x\} \leq e^{-x}.
\end{eqnarray}
\done

\begin{center}
    {\bf Appendix B: Algorithm 1 \emph{(Numerical Calculation of
$\sigma_{I}(\theta)$ and $\rho_{I}(\theta)$)}}
\end{center}

Let $y(t)= \sup_{s\geq 0}\{\frac{1}{\theta}\log{\tE e^{\theta
I(s,s+t)}}\}$. Obviously, $y(t)$ is an increasing function of $t$
with $y(0)=0$. We define axes $t$ and axes $t_\bot$ (vertical to
$t$) on a plane and we can imagine plotting $y(t)$ on it. We define
the slope of $y(t)$, $s(t) = y(t)-y(t-1)$.

We calculate $s(t)$ for $t = 1, 2, 3,...$ until it converges at some
$t^*$, i.e., $(1-\epsilon) \cdot s(t^*-1) \leq s(t^*) \leq
(1+\epsilon) \cdot s(t^*-1)$ where $\epsilon$ is a small number,
e.g. $10^{-5}$.

We draw a straight line $l(t)$ with the slope $s(t^*)$ crossing the
point $\big(t^*, y(t^*)\big)$ on the axes of $t$ and $t_\bot$.
Obviously, the line crosses the point $\big(0,
y(t^*)-s(t^*)t^*\big)$. The maximum vertical distance between $y(t)$
and $l(t)$, $v_m=\max_{0\leq t\leq t^*}\{y(t)-l(t)\}$. We shift
$l(t)$ by $v_m$ and get $\tilde {l}(t)$. Clearly, $y(t) \leq \tilde
{l}(t)$. By Definition~\ref{def:sigma_rho}, $\rho(\theta)=s(t^*)$
and $\sigma(\theta) =y(t^*)-s(t^*)t^*+v_m$.

\textbf{Yue Wang's Biography} \ \\
Yue Wang received his B.Sc degree in Mechanics and M.Sc degree in
Computer Science from Peking University, and received his PhD degree
in The Chinese University of Hong Kong. He is currently a lecturer
with School of Information, Central University of Finance and
Economics in China. His research interests lie in the performance
evaluation and optimization of wireless networks. He has served on
the TPC of Workshop on Network Calculus (WoNeCa) in 2012. And he is
a member of ACM and CCF (China Computer Federation).



\end{document}